\documentclass{ws-procs9x6}

\begin{document}

\title{Application of the Generalized Alignment Index (GALI) method
to the dynamics of multi--dimensional symplectic maps}

\author{T. MANOS$^{\mathrm{a},\mathrm{b},*}$,
Ch. SKOKOS$^{\mathrm{c}}$ and T. BOUNTIS$^{\mathrm{a}}$}

\address{$^a$Center for Research and Applications of Nonlinear Systems
(CRANS), Department of Mathematics, University of Patras, GR--26500,
Patras Greece.\\ $^b$Observatoire Astronomique de Marseille-Provence
(OAMP), 2 Place Le Verrier, 13248, Marseille, France.\\ $^c$Astronomie
et Syst\`{e}mes Dynamiques, IMCCE, Observatoire de Paris, 77
Av. Denfert--Rochereau, F--75014, Paris, France.\\ $^*$E-mail:
thanosm@master.math.upatras.gr (T. Manos)}

\begin{abstract}
We study the phase space dynamics of multi--dimensional symplectic
maps, using the method of the Generalized Alignment Index (GALI). In
particular, we investigate the behavior of the GALI for a system of
$N=3$ coupled standard maps and show that it provides an efficient
criterion for rapidly distinguishing between regular and chaotic
motion.
\end{abstract}


\bodymatter

\section{Introduction}\label{intro}
The distinction between regular and chaotic motion in conservative
dynamical systems is fundamental in many areas of applied sciences.
This distinction is particularly difficult in systems with many
degrees of freedom, basically because it is not feasible to
visualize their phase space. Thus, we need fast and accurate tools
to obtain information about the chaotic vs. regular nature of the
orbits of such systems and characterize efficiently large domains in
their phase space as ordered, chaotic, or ``sticky" (which lie
between order and chaos).

In this paper we focus our attention on the method of the
Generalized ALignment Index (GALI), which was recently introduced
and applied successfully for the distinction between regular and
chaotic motion in Hamiltonian systems \cite{sk:6}. The GALI method
is a generalization of the Smaller Alignment Index (SALI) technique
of chaos detection \cite{sk:1,sk:3,sk:5}. We present some
preliminary results of the application of GALIs on the dynamical
study of symplectic maps, considering in particular the case of a
6--dimensional (6D) system of three coupled standard maps
\cite{Kan:1}. It is important to note that maps of this type have
been extensively studied in connection with the problem of the
stability of hadron beams in high energy accelerators, see
\cite{BouSko} and references therein.

Our numerical results reported here verify the theoretically
predicted behavior of GALIs obtained in \cite{sk:6} for Hamiltonian
systems. In addition we study in more detail the behavior of chaotic
orbits which visit different regions of chaoticity in the phase
space of our system.

\section{Definition and behavior of GALI}
\label{GALI_def}

Let us first briefly recall the definition of GALI and its behavior
for regular and chaotic motion, adjusting the results obtained in
\cite{sk:6} to symplectic maps. Considering a 2$N$--dimensional map,
we follow the evolution of an orbit (using the equations of the map)
together with $k$ initially linearly independent deviation vectors
of this orbit
$\overrightarrow{\nu}_{1},\overrightarrow{\nu}_{2},...,
\overrightarrow{\nu}_{k}$ with $2\leq k \leq 2N$ (using the
equations of the tangent map).  The Generalized ALignment Index of
order $k$ is defined as the norm of the wedge or exterior product of
the $k$ unit deviation vectors:
\begin{equation}\label{GALI:0}
    GALI_{k}(n)=\parallel \hat{\nu}_{1}(n)\wedge \hat{\nu}_{2}(n)
    \wedge ... \wedge \hat{\nu}_{k}(n) \parallel
\end{equation}
and corresponds to the volume of the generalized parallelepiped, whose
edges are these $k$ vectors. We note that the hat ($\,\hat{}\,$) over
a vector denotes that it is of unit magnitude and that $n$ is the
discrete time.

In the case of a chaotic orbit all deviation vectors tend to become
linearly dependent, aligning in the direction of the eigenvector
which corresponds to the maximal Lyapunov exponent and GALI$_{k}$
tends to zero exponentially following the law \cite{sk:6}:
\begin{equation}\label{GALI:1}
GALI_{k}(n)\propto
e^{-[(\sigma_{1}-\sigma_{2})+(\sigma_{1}-\sigma_{3})+...+(\sigma_{1}-\sigma_{k})]n},
\end{equation}
where $\sigma_1, \ldots, \sigma_k$ are approximations of the first
$k$ largest Lyapunov exponents. In the case of regular motion on the
other hand, all deviation vectors tend to fall on the
$N$--dimensional tangent space of the torus on which the motion
lies. Thus, if we start with $k\leq N$ general deviation vectors
they will remain linearly independent on the $N$--dimensional
tangent space of the torus, since there is no particular reason for
them to become aligned. As a consequence GALI$_{k}$ remains
practically constant for $k\leq N$. On the other hand, GALI$_{k}$
tends to zero for $k>N$, since some deviation vectors will
eventually become linearly dependent, following a power law which
depends on the dimensionality of the torus on which the motion lies
and on the number $m$ ($m \leq N$ and $m \leq k$) of deviation
vectors initially tangent to this torus. So, the behavior of
GALI$_k$ for regular orbits is given by \cite{sk:6,ChrisBou}
\begin{equation}\label{GALI:2}
    GALI_{k}(n)\propto \left\{
                         \begin{array}{ll}
                            constant & \hbox{if $2\leq k \leq N$}\\
                           \frac{1}{n^{2(k-N)-m}} & \hbox{if $N<k\leq
                           2N$ and $0 \leq m < k-N$} \\
                           \frac{1}{n^{k-N}} & \hbox{if $N<k\leq 2N$
                           and $m \geq k-N$}
                         \end{array} .
                       \right.
\end{equation}

\section{Dynamical study of a 6D standard map}
As a model for our study we consider the 6D map:
\begin{equation}\label{6Dmap}
\begin{array}{lll}
  x_{1}' &=& x_{1} + x_{2}' \\ x_{2}' &=& x_{2} +
  \frac{K}{2\pi}\sin(2\pi x_{1}) -
  \frac{\beta}{2\pi}\{\sin[2\pi(x_{5}-x_{1})]+\sin[2\pi(x_{3}-x_{1})]\}\\
  x_{3}' &=& x_{3} + x_{4}' \\ x_{4}' &=& x_{4} +
  \frac{K}{2\pi}\sin(2\pi x_{3}) -
  \frac{\beta}{2\pi}\{\sin[2\pi(x_{1}-x_{3})]+\sin[2\pi(x_{5}-x_{3})]\}\\
  x_{5}' &=& x_{5} + x_{6}' \\ x_{6}' &=& x_{6} +
  \frac{K}{2\pi}\sin(2\pi x_{5}) -
  \frac{\beta}{2\pi}\{\sin[2\pi(x_{1}-x_{5})]+\sin[2\pi(x_{3}-x_{5})]\}
  \end{array}
\end{equation}
which consists of three coupled standard maps \cite{Kan:1} and is a
typical nonlinear system, in which regions of chaotic and
quasi--periodic dynamics are found to coexist. Note that each
coordinate is given modulo 1 and that in our study we fix the
parameters of the map (\ref{6Dmap}) to $K=3$ and $\beta=0.1$.

\begin{figure*}
{\hspace{-0.5cm}
\includegraphics[height=0.35\textheight]{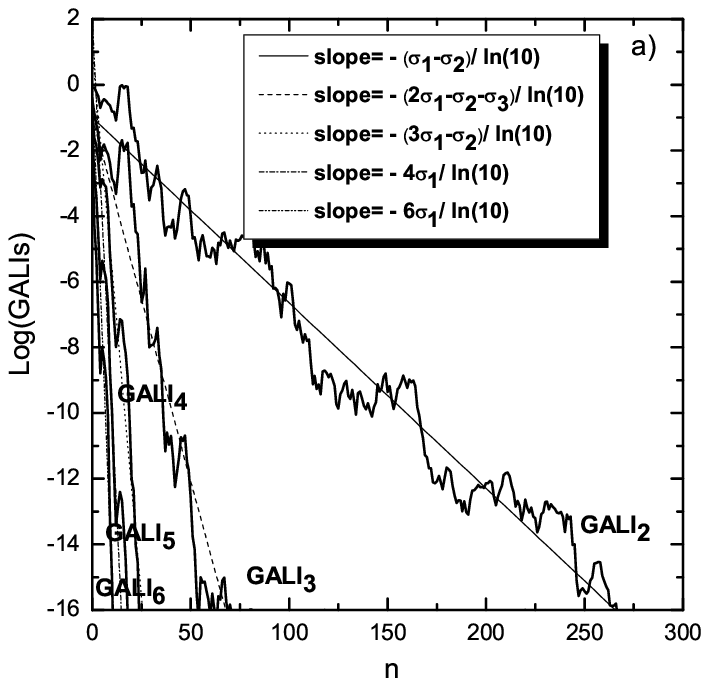}}
\hspace{-1.cm}
{\includegraphics[height=0.35\textheight]{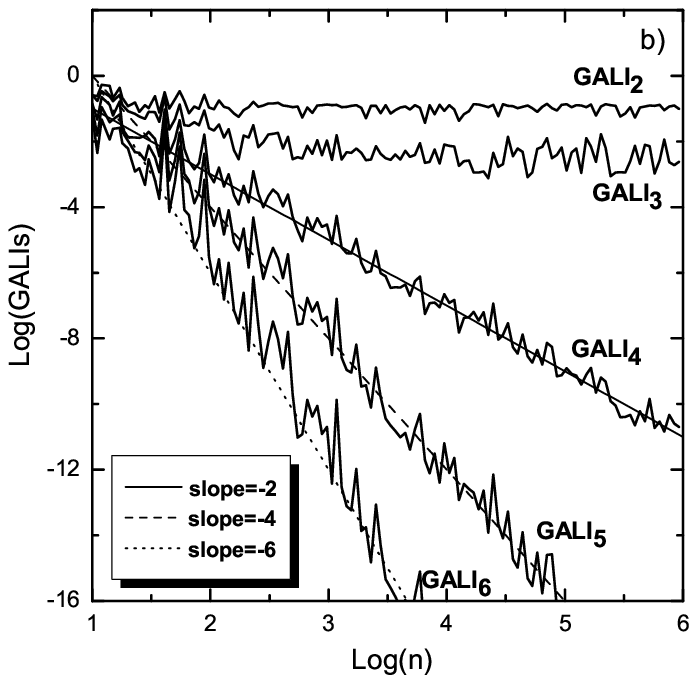}
\hspace{-.5cm}} \caption{The evolution of GALI$_k$, $k=2,\ldots, 6$,
with respect to the number of iteration $n$ for a) the chaotic orbit
C1 and b) the regular orbit R1. The plotted lines correspond to
functions proportional to $ e^{-(\sigma_1 - \sigma_2)n}$, $e^{-(2
\sigma_1 - \sigma_2-\sigma_3)n} $, $ e^{-(3 \sigma_1 - \sigma_2)n}
$, $e^{-4\sigma_1 n} $, $e^{-6\sigma_1 n} $ for $\sigma_{1}= 0.70$,
$\sigma_{2}= 0.57$, $\sigma_{3}= 0.32$ in a) and proportional to
$n^{-2}$, $n^{-4}$, $n^{-6}$ in b).} \label{6Dsm:1}
\end{figure*}

In order to verify numerically the validity of equations
(\ref{GALI:1}) and (\ref{GALI:2}), we shall consider two typical
orbits of map (\ref{6Dmap}), a chaotic one with initial condition
$x_{1}=x_{3}=x_{5}= 0.8$, $x_{2}=0.05$, $x_4=0.21$, $x_6=0.01$
(orbit C1) and a regular one with initial condition
$x_{1}=x_{3}=x_{5}= 0.55$, $x_{2}=0.05$, $x_4=0.01$, $x_6=0$ (orbit
R1). In figure \ref{6Dsm:1} we see the evolution of GALI$_k$,
$k=2,\ldots, 6$, for these two orbits.

It is well--known that in the case of symplectic maps the Lyapunov
exponents are ordered in pairs of opposite signs \cite{LL92}. Thus,
for a chaotic orbit of the 6D map (\ref{6Dmap}) we have
$\sigma_1=-\sigma_6$, $\sigma_2=-\sigma_5$, $\sigma_3=-\sigma_4$ with
$\sigma_1 \geq \sigma_2 \geq \sigma_3 \geq 0$. So for the evolution of
GALI$_k$, equation (\ref{GALI:1}) gives
\begin{equation}  \begin{array}{c}
\mbox{GALI}_2(n) \propto e^{-(\sigma_1 - \sigma_2)n},\,\,\,
\mbox{GALI}_3(n) \propto e^{-(2 \sigma_1 - \sigma_2-\sigma_3)n}, \\
\mbox{GALI}_4(n) \propto e^{-(3 \sigma_1 - \sigma_2)n}, \,\,\,
\mbox{GALI}_5(n) \propto e^{-4\sigma_1 n},\,\,\, \mbox{GALI}_6(n)
\propto e^{-6\sigma_1 n}.
\end{array}\label{eq:ch_approx}
\end{equation}
The positive Lyapunov exponents of the chaotic orbit C1 were found to
be $\sigma_{1}\approx 0.70$, $\sigma_{2}\approx 0.57$,
$\sigma_{3}\approx 0.32$. From the results of figure \ref{6Dsm:1}a) we
see that the functions of equation (\ref{eq:ch_approx}) for
$\sigma_{1}= 0.70$, $\sigma_{2} =0.57$, $\sigma_{3}= 0.32$ approximate
quite accurately the computed values of GALIs.

For the regular orbit R1 we first considered the general case where no
initial deviation vector is tangent to the torus where the orbit
lies. Thus, for the behavior of GALI$_k$, $k=2,\ldots, 6$, equation
(\ref{GALI:2}) yields for $m=0$
\begin{equation} \begin{array}{c}
\mbox{GALI}_2(n) \propto \mbox{constant},\,\,\, \mbox{GALI}_3(n)
\propto \mbox{constant} ,\,\,\,\mbox{GALI}_4(n) \propto
\frac{1}{n^2}, \\
\mbox{GALI}_5(n) \propto \frac{1}{n^4},\,\,\, \mbox{GALI}_6(n)
\propto \frac{1}{n^6}.
\end{array}\label{eq:re_approx}
\end{equation}
From the results of figure \ref{6Dsm:1}b) we see that the
approximations appearing in (\ref{eq:re_approx}) describe very well
the evolution of GALIs.

\begin{figure*}
\hspace{-0.4cm}
{\includegraphics[height=0.245\textheight]{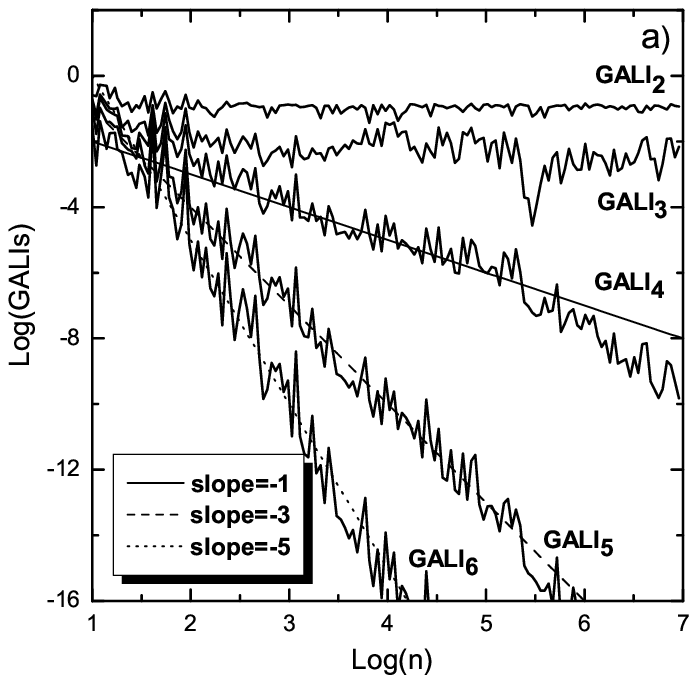}}
\hspace{-.95cm}
{\includegraphics[height=0.245\textheight]{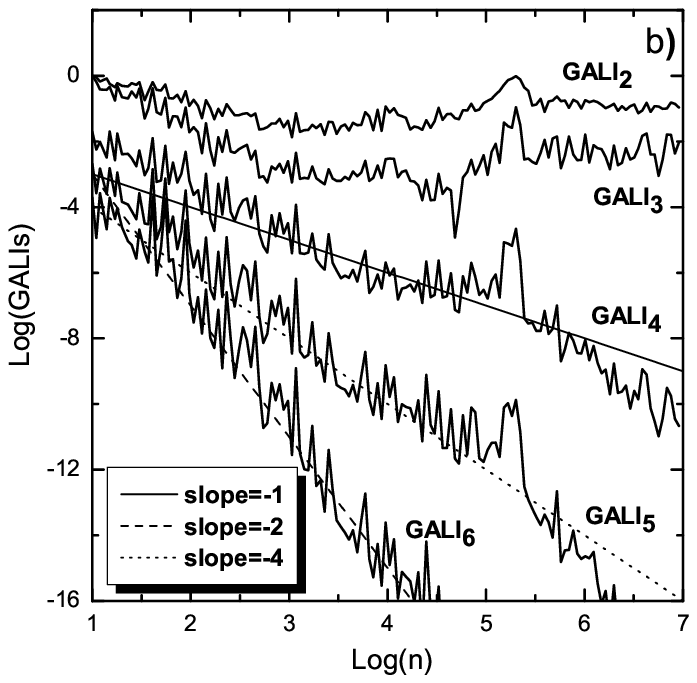}}
\hspace{-0.95cm}
{\includegraphics[height=0.245\textheight]{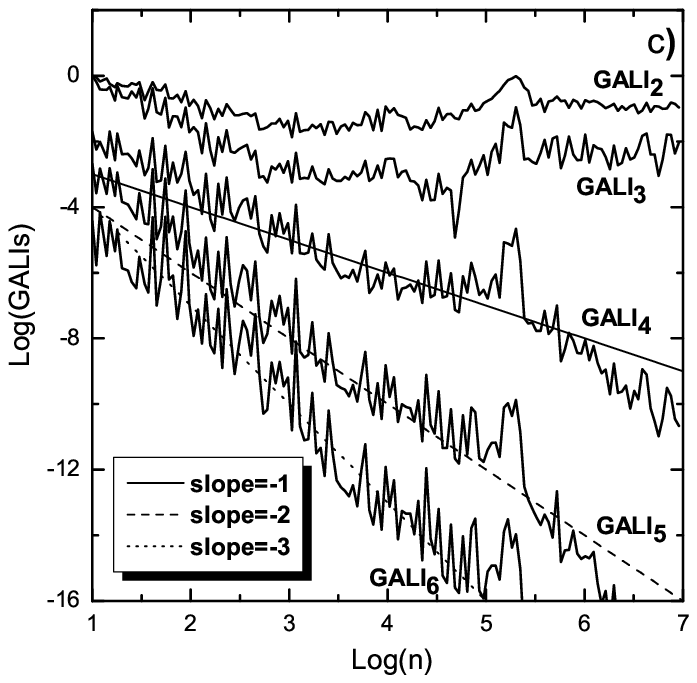}}
\hspace{-0.5cm} \caption{Evolution of GALI$_k$, $k=2,\ldots, 6$, for
the regular orbit R1 on a log--log scale, for different values of
the number $m$ of deviation vectors initially tangent on the torus
on which the motion occurs: a) $m=1$, b) $m=2$, c) $m=3$. In every
panel lines corresponding to particular power laws are also
plotted.} \label{6D_sm_dev.vect:2}
\end{figure*}

In order to verify the validity of equation (\ref{GALI:2}) for $1
\leq m \leq 3$, in the case of regular motion, we evolve orbit R1
and three random initial deviation vectors for a large number of
iterations (in our case for $5\times 10^{7}$ iterations), in order
for the three deviation vectors to fall on the tangent space of the
torus. Considering the current coordinates of the orbit as initial
conditions and using $m=1$ or $m=2$ or $m=3$ of these vectors (that
lie on the tangent space of the torus) as initial deviation vectors
we start the computation of GALIs' evolution. We note that the rest
$6-m$ initial deviation vectors needed for our computation are
randomly generated so that they do not lie on the tangent space of
the torus. The results of these calculations are presented in figure
\ref{6D_sm_dev.vect:2}, where the evolution of GALI$_k$,
$k=2,\ldots, 6$, for different values of $m$ is plotted. Figure
\ref{6D_sm_dev.vect:2} clearly illustrates that equation
(\ref{GALI:2}) describes accurately the behavior of GALIs for
regular motion also in the case where some of the initial deviation
vectors are chosen in the tangent space of the torus.

\begin{figure}
\hspace{-0.4cm}
{\includegraphics[height=0.235\textheight]{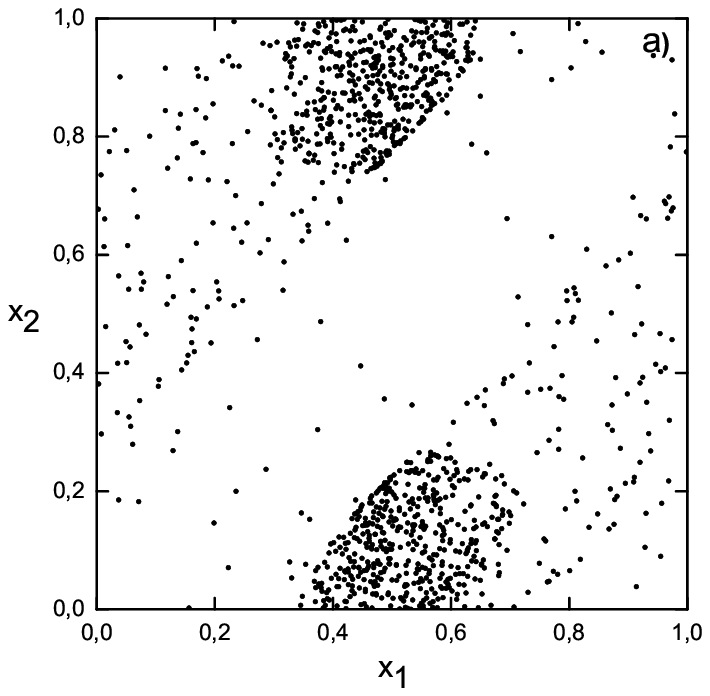}}
\hspace{-.95cm}
{\includegraphics[height=0.245\textheight]{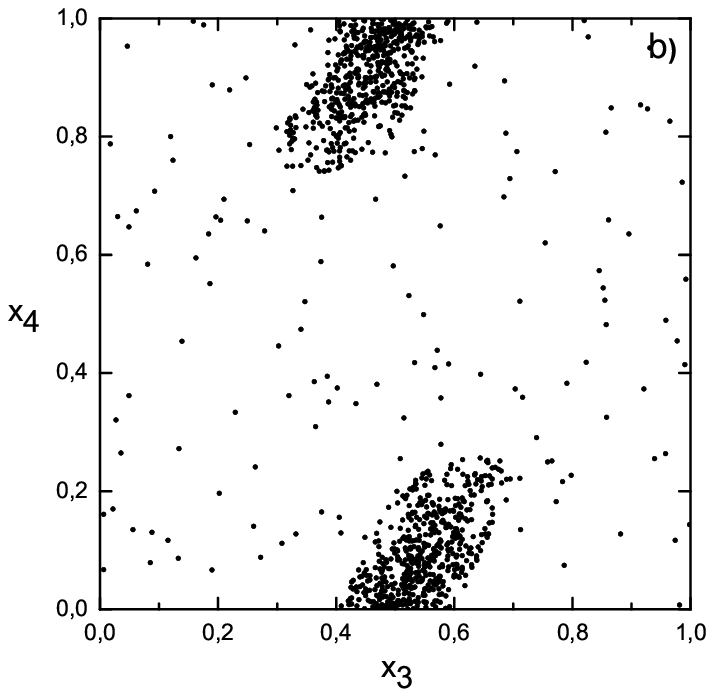}}
\hspace{-.95cm}
{\includegraphics[height=0.245\textheight]{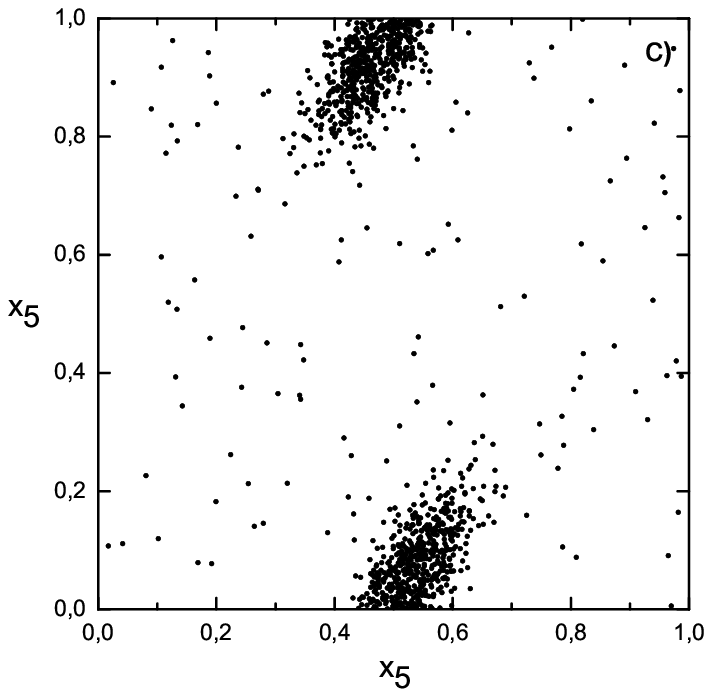}}
\hspace{-0.5cm} \caption{Projections on the planes a)
$(x_{1},x_{2})$, b) $(x_{3},x_{4})$ and c) $(x_{5},x_{6})$ of the
first $1000$ successive points of the chaotic orbit C2.}
\label{6Dsm_pss_dif_region:1}
\end{figure}

Let us now consider the case of a chaotic orbit which visits different
regions of chaoticity in the phase space of the map. The orbit with
initial conditions $x_{1}=x_{3}=x_{5}= 0.55$, $x_{2}=0.05$,
$x_4=0.21$, $x_6=0.0$ (orbit C2) exhibits this behavior as can be seen
from the projections of its first $1000$ successive consequents on
different 2--dimensional planes plotted in figure
\ref{6Dsm_pss_dif_region:1}.

\begin{figure}
\hspace{-0.4cm}
{\includegraphics[height=0.235\textheight]{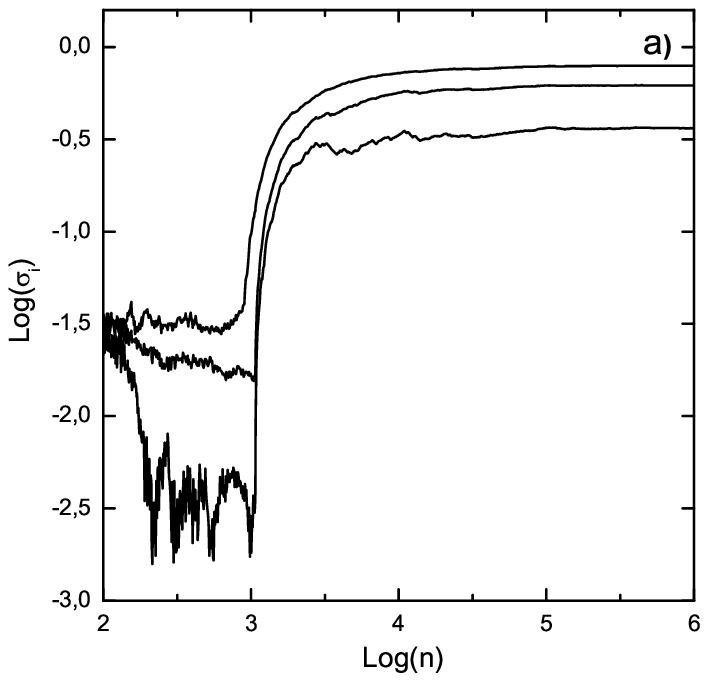}}
\hspace{-.95cm}
{\includegraphics[height=0.224\textheight]{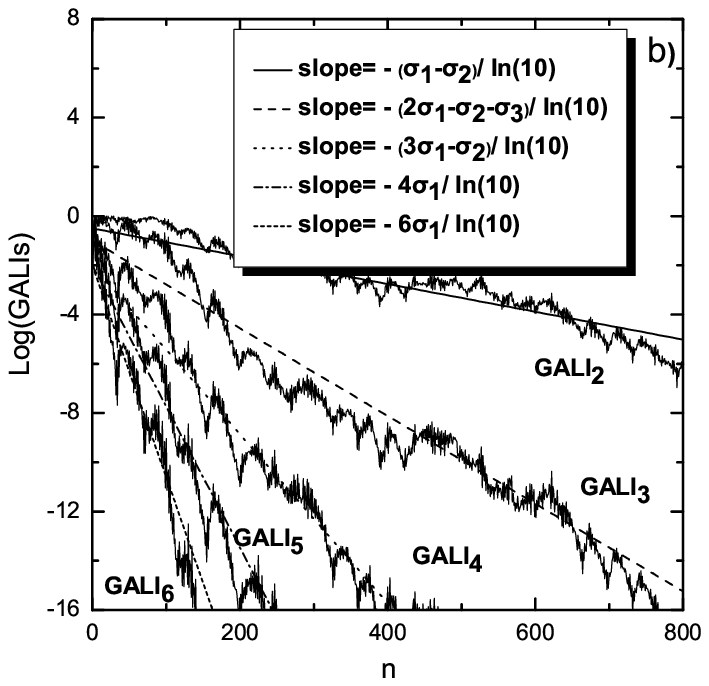}}
\hspace{-.95cm}
{\includegraphics[height=0.235\textheight]{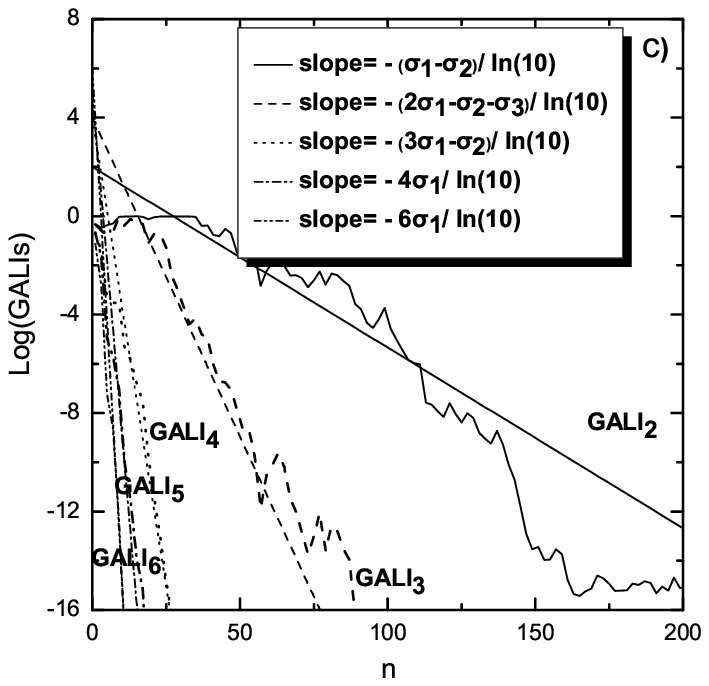}}
\hspace{-0.5cm}

\caption{a) The evolution of the three positive Lyapunov exponents
of the chaotic orbit C2. The evolution of GALI$_k$, $k=2,\ldots,6$,
with respect to the number of iteration $n$ for the same orbit when
we use as initial condition of the orbit its coordinates at b) $n=0$
and at c) $n=10^6$ iterations. The plotted lines in b) and c)
correspond to functions proportional to $e^{-(\sigma_1 -
\sigma_2)n}$, $e^{-(2 \sigma_1 - \sigma_2-\sigma_3)n}$, $e^{-(3
\sigma_1 - \sigma_2)n}$, $e^{-4\sigma_1 n}$, $e^{-6\sigma_1 n}$ for
$\sigma_{1}= 0.033$, $\sigma_{2}= 0.02$, $\sigma_{3}= 0.005$ in b)
and $\sigma_{1}= 0.793$, $\sigma_{2}= 0.624$, $\sigma_{3}= 0.365$ in
c).} \label{6D_Lyap_GALIs:1}
\end{figure}

The projections look erratic, indicating that the orbit is chaotic.
However, we also observe in all three projections that the C2 stays
`trapped' for many iterations in two oval--shaped regions and
eventually escapes entering the big chaotic sea around these
regions. This behavior is also depicted in the evolution of the
Lyapunov exponents of the orbit (see figure \ref{6D_Lyap_GALIs:1}a).
The three positive Lyapunov exponents are seen to fluctuate around
$\sigma_{1}\approx 0.033$, $\sigma_{2}\approx 0.02$,
$\sigma_{3}\approx 0.005$ for about $1000$ iterations exhibiting
`jumps' to higher values when the orbit enters the big chaotic sea,
stabilizing around $\sigma_{1}\approx 0.793$, $\sigma_{2}\approx
0.624$, $\sigma_{3}\approx 0.365$.

Let us now study how the GALIs are influenced by the fact that orbit
C2 visits two different regions of chaoticity characterized by
different values of Lyapunov exponents. Since C2 is a chaotic orbit,
its GALIs should tend exponentially to zero following the laws of
equation (\ref{eq:ch_approx}). Thus, starting the computation of the
GALI$_k$, $k=2,\ldots, 6$, from an initial point of C2 located in
the first chaotic sea, we see that the slopes of the exponential
decay of GALIs are well described by equation (\ref{eq:ch_approx})
using for $\sigma_1$, $\sigma_2$, $\sigma_3$ the approximate values
of the Lyapunov exponents of the small chaotic region (figure
\ref{6D_Lyap_GALIs:1}b). On the other hand, using as initial
condition for this chaotic orbit its coordinates after $10^6$
iterations, when the orbit has escaped in the second chaotic region,
the evolution of GALIs is again well approximated by equation
(\ref{eq:ch_approx})
 but this time for $\sigma_{1}= 0.793$, $\sigma_{2}=
0.624$, $\sigma_{3}= 0.365$, which are the approximations of the
Lyapunov exponents of the big chaotic sea (figure
\ref{6D_Lyap_GALIs:1}c). Thus, we see that in this case also the
$\sigma_i$, $i=1,\ldots, k$ which appear in equation
(\ref{eq:ch_approx}) are good approximations of the first $k$
Lyapunov exponents of the large chaotic region in which the orbit
eventually wanders after about $10^3$ iterations.

\section{Conclusions}

In this paper we verified the theoretically predicted behavior of
the Generalized Alignment Index (GALI) by considering some
particular regular and chaotic orbits of a $2N$--dimensional
symplectic map with $N=3$. In particular, we showed numerically that
all GALI$_k$, $2\leq k\leq 2N$, tend to zero exponentially for
chaotic orbits, while for regular orbits they remain different from
zero for $2\leq k\leq N$ and tend to zero, following particular
power laws, for $N < k\leq 2N$. Thus, by using GALI$_k$ with
sufficiently large $k$, one can infer quickly the nature of the
dynamics much faster than it is possible by using other methods.

Also, the study of chaotic orbits which visit different regions of
chaoticity in the phase space of the system, provides further
evidence that the exponents of the exponential decay of GALIs are
related to the local values of Lyapunov exponents. Thus, we have
shown that the different behaviors of the GALIs for regular and
chaotic orbits can be used for the fast and accurate identification
of regions of chaoticity and regularity in the phase space of
symplectic maps. We plan to investigate this further in a future
publication concerning maps with $N>>3$, which describe arrays of
conservative nonlinear oscillators.

\section{Acknowledgments}
T.~Manos was partially supported by the ``Karatheodory" graduate
student fellowship No B395 of the University of Patras, the program
``Pythagoras II" and the Marie Curie fellowship No
HPMT-CT-2001-00338. Ch.~Skokos was supported by the Marie Curie
Intra--European Fellowship No MEIF--CT--2006--025678. The first author
(T.~M.) would also like to express his gratitude to the Institut de
M\'{e}canique C\'{e}leste et de Calcul des Eph\'{e}m\'{e}rides (IMCCE)
of the Observatoire de Paris for its excellent hospitality during his
visit in June 2006, when part of this work was completed.

\end{document}